# Ensuring high QoE for DASH-based clients using deterministic network calculus in SDN networks

Oussama El Marai*, Jonathan Prados-Garzon*, Miloud Bagaa*, and Tarik Taleb*§
oussama.elmarai@aalto.fi, jonathan.prados-garzon@aalto.fi, miloud.bagaa@aalto.fi, tarik.taleb@aalto.fi
*Aalto University, Espoo, Finland.
§University of Oulu, 90570 Oulu, Finland.

*Abstract*—HTTP Adaptive Streaming (HAS) is becoming the de-facto video delivery technology over best-effort networks nowadays, thanks to the myriad advantages it brings. However, many studies have shown that HAS suffers from many Quality of Experience (QoE)-related issues in the presence of competing players. This is mainly caused by the selfishness of the players resulting from the decentralized intelligence given to the player. Another limitation is the bottleneck link that could happen at any time during the streaming session and anywhere in the network. These issues may result in wobbling bandwidth perception by the players and could lead to missing the deadline for chunk downloads, which result in the most annoying issue consisting of rebuffering events. In this paper, we leverage the SoftwareDefined Networking paradigm to take advantage of the global view of the network and its powerful intelligence that allows reacting to the network changing conditions. Ultimately, we aim at preventing the re-buffering events, resulting from deadline misses, and ensuring high QoE for the accepted clients in the system. To this end, we use Deterministic Network Calculus (DNC) to guarantee a maximum delay for the download of the video chunks while maximizing the perceived video quality. Simulation results show that the proposed solution ensures high efficiency for the accepted clients without any rebuffering events which result in high user QoE. Consequently, it might be highly useful for scenarios where video chunks should be strictly downloaded on-time or ensuring low delay with high user QoE such as serving video premium subscribers or remote control/driving of an autonomous vehicle in future 5G mobile networks.

## I. Introduction

Internet video traffic is already overwhelming current network's traffic mainly carried out over 4G and is expected to reach 82% by 2022 [1]. The massive proliferation of different kind of mobile devices and the faster and ubiquitous connectivity offered by 4G networks have played a key role in video traffic dominance. By 2022, mobile devices are expected to be the most popular to watch video [1]. Additionally, the faster connectivity expected by 5G will essentially contribute to increasing the users' engagement for video watching which will result in almost doubling the amount of carried data over the network. As a result, a huge pressure will be exerted on the underlying infrastructure which adversely impacts the user's Quality of Experience (QoE) [2].

Ensuring high user's QoE in Over-the-top (OTT) streaming services is a challenging process. It requires involving the different actors of an HAS-based system, namely the client, server and network. Furthermore, intelligence should be injected at each actor to give it in one hand the autonomy to react to the changing conditions, and on the other hand, allowing cooperation and interaction between the different actors by inter-changing crucial information among themselves in order to efficiently utilize the resources and achieve user satisfaction.

The first actor in a streaming system that we want to add intelligence is the player or the consumer of the video stream. This intelligence should give it the ability to adapt to the network conditions and select adequate video quality in a way to avoid video stalls and maximize its QoE. One highly interesting technology that enables such intelligence is HAS. The latter is a new paradigm that has shown its efficiency in delivering video content on a large scale. It rapidly becomes the predominant video delivery technique in the contemporary web, thanks to its myriad advantages among which the adaptability to network conditions, the large-scale deployment, and the cost-effectiveness by re-using existing infrastructure. This approach sounds interesting and enable high flexibility, but entails selfishness since each player independently tries to increase its representation to get higher QoE. This philosophy generates unwanted behavior, such as oscillations, that harms the overall performance of the system in a multi-players scheme.

The second actor in a streaming system is the server. In HAS-based system, the server's role is very simple and limited. It consists of handling the client's HTTP requests and supplying the requested content in HTTP responses. This allows for better scalability and easier deployment. However, important information is available at the server, such as the number of connected clients and the server's bandwidth, that could be helpful to achieve higher system performance. Hence, injecting intelligence at the server [3], [4] is highly important for the overall outcome.

The last actor consists of the network through which the video traffic is carried out. In fact, it is the most important actor that has a direct impact on the client's decision. Given the limited resources of the network, even though considering the great advances in the network bandwidth provided by 4G and 5G networks, and the increasing demand for video traffic exerting high pressure, the need to incorporate intelligence to the network becomes a must with the ultimate objective of providing higher QoS. Software-Defined Networking (SDN) is a new paradigm that enables such intelligence through OpenFlow protocol [5]–[9]. It consists of decoupling the control plane from the data plane in computer networks. By doing so, Internet Service Providers (ISPs) can programmatically change their routers' policies that influence the traffic direction through the controller. Depending on the situation in the network and the policy of the ISP, the controller makes decisions on how to route the traffic in the network by pushing rules to the switches. This approach enables flexible and intelligent management of the network resources which result in improving the overall system performance and achieve higher user satisfaction.

In this paper, we are more interested in the last actor (i.e. network) to ensure a stall-free session for each player in SDNenabled networks. To this aim, we have used Deterministic Network Calculus (DNC) [10] theory to compute the maximum delay to download a chunk from a given representation over the different paths in the network linking the client to the server. Our approach allows selecting for each client the best path in the network that ensures the highest possible video quality having a maximum delay lower than the segment length.

The remainder of this paper is structured as follows. In Section II we will present the related work. Section III describes our system model and problem formulation. In Section IV, we provide a detailed description of the DNC model and the proposed solution. Simulation results are presented in Section V. Finally, Section VI summarizes our findings and highlights future research directions.

## II. Related Work

The predominant HAS technology has gained an immense interest by researchers and industries in recent years and this is thanks to the interesting achieved results and its promising rosy prospects. All the proposed work can be classified into client-side, server-side or in-network solutions [11]. In this section, we are interested in the last category, more specifically to the work exploiting SDN networks for improving HAS performance.

In [12], the authors proposed the so-called SABR (SDN assisted Adaptive BitRate) architecture, that leverages SDN network, to feed the DASH clients, through REST APIs, with the relevant measurements on the network conditions, such as available bandwidth estimation and cache occupancy, to improve the clients' estimation accuracy in order to achieve better user's QoE. Bentaleb *et al.* target in [6] the scalability issue of DASH systems when multiple heterogeneous clients compete for the same bottleneck link. They propose SDNDASH architecture that leverages SDN networks for dynamic resource allocation to maximize the users' QoE. To this end, they implement a separate resource management application, external to the SDN controller, that communicates with both DASH client and server through REST APIs. This application contains different modules and components, each of them is responsible for a specific task such as data inspection and collection, bandwidth estimation and recording network/server statistics. Another resource management application is implemented inside the controller. It is responsible for the effective resource allocation and monitoring, QoS provisioning, and allows the deployment of the selected QoS policy via OpenFlow messages. An important module in the external application, namely communication component, is responsible for the mapping of the QoE policy to the QoS policy and communicate it to the internal application through REST API. An optimization problem is also provided that aims at maximizing the QoE for each client under four constraints: buffer occupancy level, available bandwidth, content type, and device capabilities. The model is solved using online MPC (Model Predictive Control) algorithm.

In [13], the authors studied the impact of three different network-assisted approaches, namely *Bandwidth Reservation (BR)*, *Bitrate Guidance (BG)* and a hybrid approach combining the two previous approaches, on the QoE properties. For the *Bandwidth Reservation* scheme, a bandwidth slice is assigned to the video flows and the client is responsible on the selection of the video bitrate, while in the *Bitrate Guidance* approach the video bitrate selection is performed by a centralized algorithm in the network. For each scenario, the authors formulate an optimization problem, namely *Discrete Video Quality fair assignment* and *Continuous Video Quality fair assignment*. The authors conclude that the presence of the networkassisted approaches produces a better video quality fairness compared to purely client-side algorithms. Particularly, the *Bitrate Guidance* approach offers better results compared to

*Bandwidth Reservation*. Kleinrouweler *et al.* in [8] leverage SDN networks and employ a hybrid adaptation framework where the clients are assisted by the *Service Manager* which plays the role of a proxy between the Network Controller and DASH players. The in-network *Service Manager* has a global view on the active players and the network resources, therefore it is capable of assisting the players in their decisions and performs network resources allocation, through network controller, to achieve the fair share using queuing strategy. The proposed architecture is evaluated under a fluctuate WiFi network and the results show a significant improvement in the video quality (up to a double), streaming stability and fair sharing of the bandwidth among players.

In [14], the authors proposed a traffic shaping approach that leverages SDN networks in a wireless environment. The main objectives are to improve the bandwidth utilization, by taking advantage of the long OFF periods of idling clients to increase the bandwidth of active (ON) clients, improving the QoE and reducing the mobile devices' energy consumption. To do that, the client's controller collects some network measurements, like RTT, and send them to the SDN controller which in turn computes the fair shaping rate and sends the decision back to the client. Similarly, in [15] the authors studied the impact of the network traffic shaping on the adaptive video streaming and conclude that the individual traffic shaping presents better performance in terms of maintaining a balance between the perceived video quality and fairness among clients. In [16], the authors proposed a reactive solution, when a client experience rebuffering events, that changes the delivery node in CDNs in order to assign a new one and find the best path that provides better QoE to the user. In [17], the authors proposed a multiobjective model for segment-based path selection in SDNenabled networks for the aim to provide high QoE based on the paths' available bandwidth, the segment bit-rate, and the path length. The obtained results of the proposed approach show better performances compared to the traditional besteffort routing. In order to reduce latency and outage duration, Georgopoulos *et al.* have proposed in [18] a transparent innetwork *cache as a service* approach for streaming VoD content, by leveraging SDN networks, to place the video content as close as possible to the end-user. The proposed approach helps to improve the user's QoE and efficiently use the network bandwidth by avoiding burden the network with identical content. In [19], the authors proposed the QFF (QoE Fairness Framework), an OpenFlow-assisted framework that aims to provide a user-level QoE fairness for heterogeneous clients. The proposed QFF is composed of an Orchestrating Module (OM) that takes in input statistics about the network and the clients' requests from both *Network Inspector* and *MPD Parser* modules respectively. Using the statistic information, a *Utility Function* and *Optimization Function* are employed in order to decide the QoE fairness, and then the decision is pushed to the network and communicated to DASH clients through *Flow Tables Manager* and *DASH Plugin* modules respectively.

Unlike prior works that leverage SDN networks for unicast streaming, Shen *et al.* investigate in [20] the use of multicast streaming in order to reduce the traffic in the network. They have proposed Recover-aware Steiner Tree (RST), an optimization problem that aims at minimizing both tree and recovery cost for multicast traffic. To solve RST, the algorithm RAERA (Recover Aware Edge Reduction Algorithm) has been proposed consisting of two essential phases: Tree Routing Phase and Recovery Selection Phase. The goal is to reduce the recovery cost by selecting recovery nodes close to the destinations when loss packets occur, hence diminish the retransmissions delay and bandwidth usage.

## III. Problem Formulation

Let us consider a full-duplex forwarding plane of an SDN network consisting of $K$ SDN switches. The SDN network interconnects $N-1$ HAS clients and one HTTP server (hereinafter referred to as hosts). This scenario can be modeled as an undirected graph G = (V,E), whereby V represents the set of network nodes and E represents the links between them. Two types of network nodes exist V = H ∪ S, where H = $\{h_1,h_2,..,h_N\}$ represents the set of hosts, and S = $\{s_1,s_2,..,s_K\}$ represents the set of switches. Each host is connected to only one access switch. Each link/edge $e \in$ E is characterized by its delay $\theta_e$ (e.g., propagation and processing latencies) and capacity $C_e$ (transmission rate).

At the server side, each video is encoded into $L$ video representations R = $\{r_1,r_2,..,r_L\}$, each of which is chopped into equal small chunks of $\tau$ duration length. We denote by $b_j$ the size (in bits) of a given chunk.

In this work, we will consider the problem of choosing the paths in the SDN network that maximizes the average video quality of the accepted HAS clients, while guaranteeing zero rebuffering events for them. Specifically, the optimization problem considered in this work is formally formulated as follows:

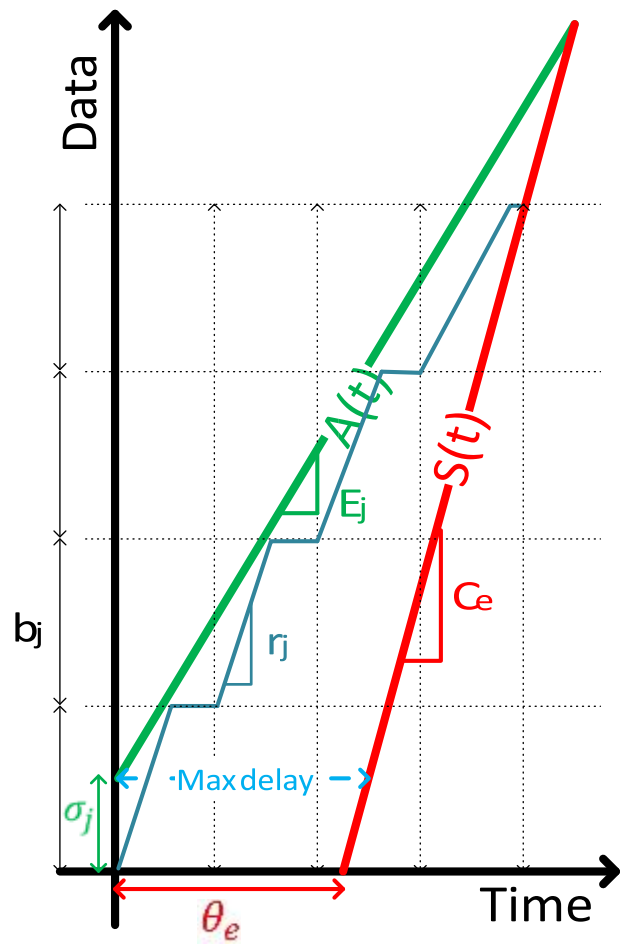

Fig. 1: ($\rho$, $\sigma$)-Upper constrained video download profile and maximum virtual delay considering a latency-rate server.

*Objective* :

$$\text{maximize} \quad \sum_{j=1}^{J} \alpha_j E_j \tag{1a}$$

*Constraint* :

$$d_j^{(chunk)}\left(u_j^{(e)}\right) \leq \frac{b_j}{E_j} = \tau \tag{1b}$$

The decision variables of the optimization problem are $u^e_j \forall j \in [1,J] \cap N$ which are binary variables indicating whether a given edge $e \in$ E is allocated to the client $j$ ($u_j^e = 1$) or not ($u_j^e = 0$). Objective (1a) aims to maximize the video quality of a set of HAS clients attached to an SDN network. Constraint (1b) assures that no HAS client is going to experience a rebuffering event. In other words, the maximum delay to download a video chunk should be lower or equal than the chunk duration $\tau$.

TABLE I: Primary notation.

| Notation | Description |
|---|---|
| $r_j$ | Maximum e2e download rate for the user $j$. |
| $\rho_j$ | Enforced average download rate for the user $j$. |
| $E_j$ | Video encoding rate or quality allocated to the user $j$. |
| $b_j$ | Maximum chunk size for the user $j$. |
| $C_e$ | Capacity of the link $e$. |
| $\theta_e$ | Delay introduced by the link $e$ regardless its workload. |
| $sp$ | The selected path. |
| $sr$ | The selected representation. |
| $\Gamma$ | List of paths between the client and the server. |
| $\Lambda$ | The bottleneck in a given path. |
| $\Phi$ | The highest bottleneck capacity. |

## IV. Performance Model

### A. Model Description

In this section we will provide a worst-case analysis of downloading a chunk in a SDN network. To that end, we employ network calculus which is a theory that relies on inequalities and min-plus algebra to estimate the performance guarantees of a network [10]. The first step in any network calculus analysis is to find an upper bound for the arrival process. For a given video quality, it is expected that the HAS server limits the data generation rate to the video encoding rate of the chosen representation. Thus, we will consider that the video download profile for the client $j$, i.e., the amount of data (for instance expressed in bits) downloaded until a given time $t$, is upper-bounded by the following function (see Fig. 1):

$$A_j(t) = \rho_j \cdot t + \sigma_j \quad (2)$$

where $\rho_j$ is the average sustainable rate and $\sigma_j$ is the burstiness. Specifically, $\rho_j = E_j$ and $\sigma_j = b_j \cdot (1 - E_j/r_j)$ (see Fig. 1). Then, the aggregated download profile of $J$ flows constrained by the above function will be, in turn, constrained by A(t) which is given by [10]:

$$A(t) = \sum_{j=1}^{J} \rho_j \cdot t + \sum_{j=1}^{J} \sigma_j \quad (3)$$

For every output port of a given SDN switch (or link), we will consider a rate-latency server model [10]. That is, its service curve, i.e., number of bits transmitted per unit of time, is lower-bounded by the following function:

$$\beta_e(t) = C_e \cdot (t - \theta_e) \quad (4)$$

The virtual maximum delay at time $t$ is upper-bounded by the horizontal distance between the arrival and the service curves [10] (see Fig. 1):

$$d(t) \leq inf\{\tau \geq 0 : A(t) \leq A^*(t+\tau)\} \quad (5)$$

$$d_{max} = \frac{b_j}{C_e} \cdot \left(1 - \frac{E_j}{r_j}\right) + \theta_e \quad (6)$$

The above result can be extended to a tandem of network elements by leveraging the blind multiplexing and concatenation properties of network calculus [10]:

- The *blind multiplexing property* provides us the service curve $\beta_1(t)$ perceived by a flow $f_1$ that is sharing the server (resource), which has a strict service curve $\beta$, with another flow $f_2$ assuming statistical multiplexing. Specifically, if $f_2$ is upper-bounded by $\alpha_2(t)$. Then, :

  $$\beta_1(t) = [\beta(t) - \alpha_2(t)]^+ \quad (7)$$

  where $[\cdot]^+$ denotes the operation $\max\{\cdot, 0\}$.
- The *concatenation property* allows us to compute an equivalent service curve for the association of a set of servers in tandem.

  $$\beta(t) = \beta_1 \otimes \beta_2 \otimes \cdots \otimes \beta_n(t) \quad (8)$$

  where the operator $\otimes$ stands for the min-plus convolution which is defined as $f \otimes g(t) = \min_{\tau \in [0,t]}\{f(\tau) + g(t - \tau)\}$.

Considering the blind multiplexing property, any incoming flow $i$ to be allocated to a given link will perceive a service curve:

$$\beta_{e,i}(t) = \left[\beta_e(t) - \sum_{j \in \mathcal{J}_e} \left(E_j \cdot t + b_j \cdot \left(1 - \frac{E_j}{r_j}\right)\right)\right]^+ \quad (9)$$

where $\mathcal{J}_e$ is the set of flows already allocated to the link $e \in \mathcal{E}$. It is easy to prove that $\beta_{e,i}(t)$ is a rate-latency service curve ($\beta_{e,i} = C_{e,i} \cdot (t - \theta_{e,i})$) with rate parameter $C_{e,i} = C_e - \sum_{j \in \mathcal{J}_e} E_j/r$, and latency parameter $\theta_{e,i} = \theta_e + \sum_{j \in \mathcal{J}_e} b_j/C_e \cdot (1 - E_j/r_j)$.

Finally, applying the concatenation property, the association of $n$ rate-latency servers is again a rate-latency server with rate parameter $C = \min_{e \in \mathcal{E}_j}\{C_e\}$ and latency $\theta = \sum_{e \in \mathcal{E}_j} \theta_e$.

Then, we can derive an end-to-end delay bound $\hat{d}_j^{(e2e)}$ for a flow $j$ traversing $n$ links.

$$\hat{d}_j^{(e2e)} = \frac{b_j}{\min_{e \in \mathcal{E}_j}\{C_e - \sum_{i \in \mathcal{J}_e} E_i\}} \cdot \left(1 - \frac{E_j}{r_j}\right) + \sum_{e \in \mathcal{E}_j} \left(\theta_e + \sum_{i \in \mathcal{J}_e, i \neq j} \frac{b_i}{C_e} \cdot \left(1 - \frac{E_i}{r_i}\right)\right) \quad (10)$$

*B. Solution*

In order to decide which video representation should be allocated to the connecting clients, we use Algorithm 1. For each newly connected client, the HAS server triggers a request to the decision server in order to find the best possible path that maximizes the video quality of the corresponding client. To do that, the decision server gets from the SDN controller the actual state of the network. Then, it browses all possible paths between the client and the server and chooses the one $sp$ with the highest bottleneck capacity of $\Phi$. The latter value is then

```
Algorithm 1: Maximum Delay Video Quality Allocation
foreach c ∈C do
    sp ← []; sr ← 0; Φ ← 0;
    Γ ← getListPaths ()
    foreach p ∈ Γ do
        Λ ← getPathBottleneck () ;   /* returns
          the lowest available capacity in
          all edges in this path*/
        if Λ > Φ then
            Φ ← Λ;
            sp ← p ;
        end
    end
    sr ← getHighestVideoQuality (Φ) ; /* returns
      the highest video quality lower
      than Φ */
    if sq =0 then
        Reject (c);
    else
        Φ ← 0;
        foreach r ∈ reversed (R ) do
            if r ≤ sr then
                Φ ← getMaxDelay (sp,sr )
                if Φ < = τ then
                    HasImpact =
                      HasImpactOnOtherClients ();
                    if HasImpact = False then
                        Accept (c,sp,r );
                        break;
                    end
                end
            end
        end
        if Φ=0 then
            Reject (c);
        end
    end
end
```

used to determine the highest possible video representation *sr* lower than $\Phi$. The next step is to loop through the video representations lower than or equal to *sr*, from the highest to the lowest, and calculate the maximum delay, using (10), for this specific path *sp*. The video representation allocated to the client is the one meeting two conditions: *i*) the maximum end-to-end delay to transmit a chunk of this representation is lower than $\tau$, *ii*) it does not prevent the video cross-flows from fulfilling their delay constraints (refer to (1b)). The cross-flows are those passing through at least one link of the selected path *sp* for the ongoing flow allocation. If all video representations do not meet the aforementioned conditions, the client is then rejected. The global architecture of the proposed system is depicted in Fig. 2.

## V. Simulation Results

We assessed the performances of our solution by means of simulations and compared it to a fair share solution. The latter selects the shortest path between the client and the server, and fairly shares all the edges in the selected path among the competing flows. The fair share solution starts rejecting the incoming flows whenever its allocation yields an encoding rate below the lowest video representation. The following performance metrics are considered in the evaluation: *i*) the average video quality of the accepted clients; *ii*) the rejection probability; and *iii*) the accumulated rebuffering time of each client.

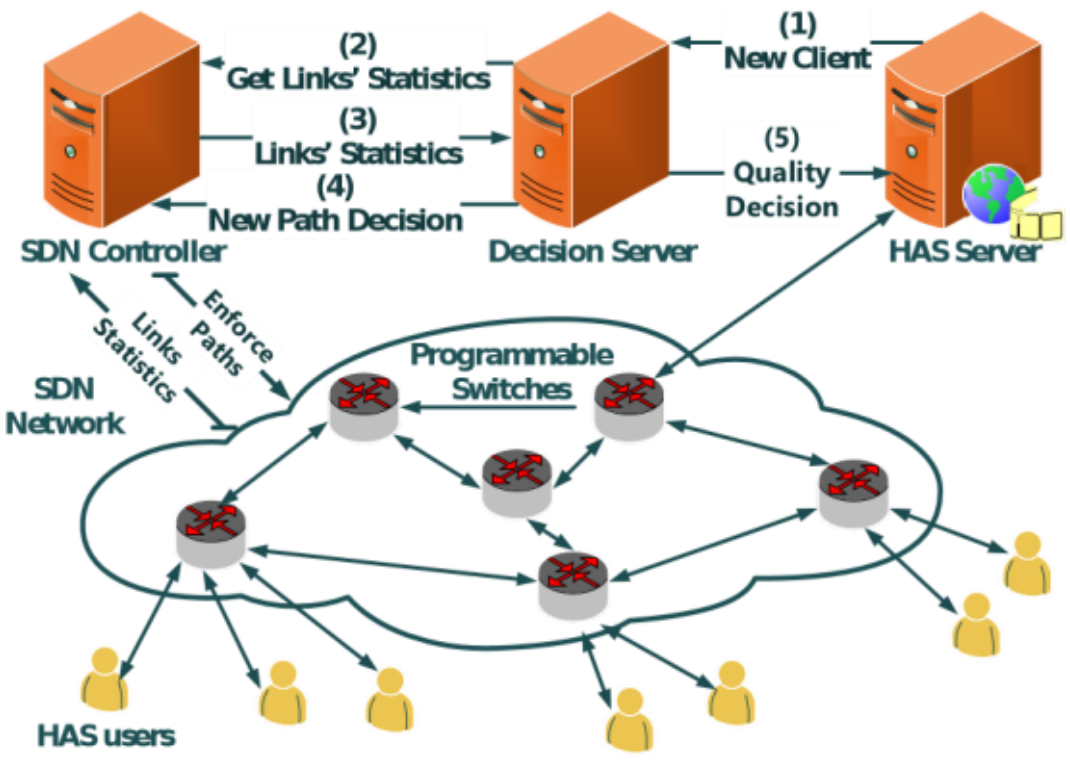

Fig. 2: System Architecture.

### A. Simulation Setup

Fig. 3 illustrates the simulated network used during the evaluation, whereby the access switches (i.e., edge switches that interconnect the clients with the rest of the network) are in red. The capacities of the last mile, links interconnecting switches and the server access link are 10 Mbps, 500 Mbps and 1000 Mbps, respectively. In each simulation, we have run 1000 clients (simulation stop condition). We used a Poisson process to simulate the client arrivals to the network. The distribution considered for the video duration is that one measured and reported in [21] whose mean is 231 seconds. In our study, the average number of clients simultaneously connected to the network varies from 20 to 200. Due to space limitation, we will show only the simulation dynamics for a scenario of 160 clients in the system. The video qualities considered are 1Mbps, 2Mbps, 3Mbps, 4Mbps, 5Mbps, and the chunk duration $\tau$ = 1*s*. To estimate the rebuffering times, we have used (10).

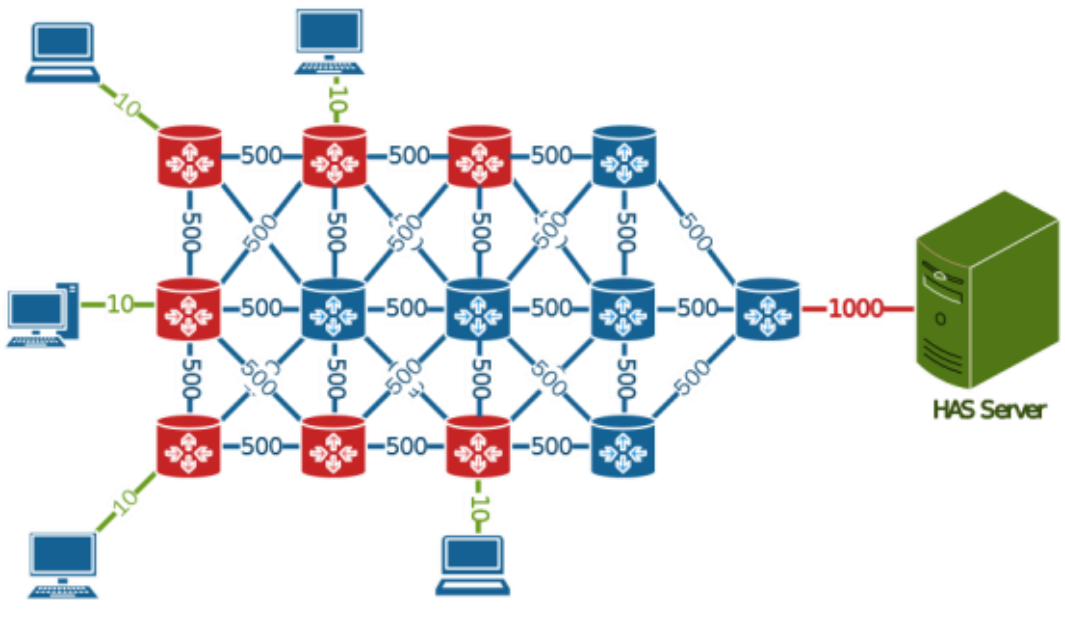

Fig. 3: Simulation Setup.

### B. Simulation Results

*1) Average Video Quality:* Fig. 4 shows the moving average of the allocated video quality when 160 clients are simultaneously connected to the server. In contrast to the fair share solution, we vividly see that our proposed solution allocates high-quality video representations to the accepted clients. Unlike the fair share solution, our proposed solution rejects clients when their admission adversely impacts the previously accepted clients. This is due to the fact that the admission control of our solution is more strict than the fair share one in order to guarantee zero rebuffering events.

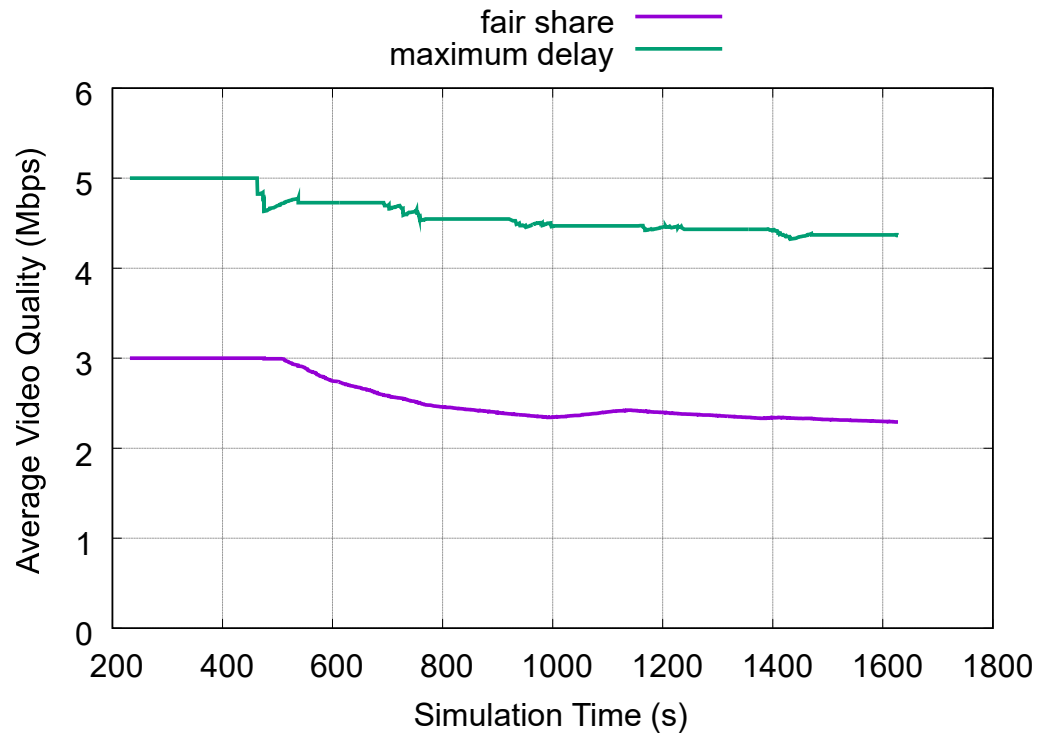

Fig. 4: Average quality for 160 incoming clients' arrival rate.

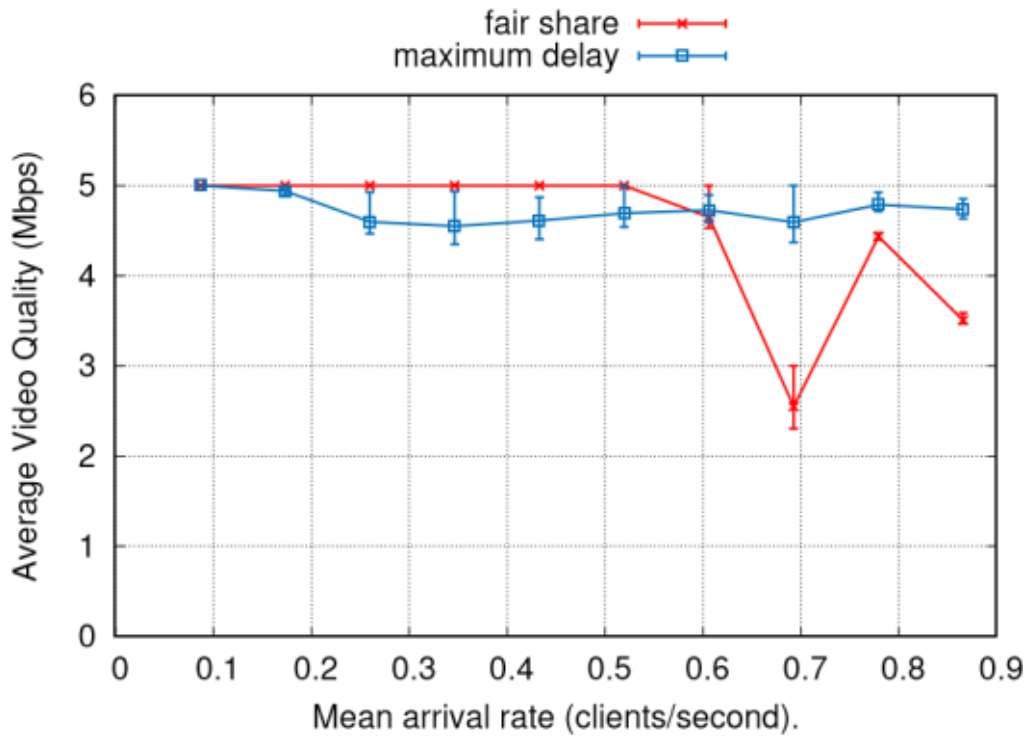

Fig. 5: Overall average video quality for accepted clients.

Fig. 5 depicts the mean as well as the 5 and 95 percentiles of the video qualities versus the average number of simultaneously connected clients. As it is observed, both solutions ensure high average video bitrates for the accepted clients at low workloads. We also note that the fair share slightly outperforms our proposed solution at low workloads, while it degrades at high workloads. However, our solution keeps allocating high video qualities regardless of the workload, thus ensuring the high QoE always.

*2) Rejection Probability:* Fig. 6 shows the rejection probability for both solutions at each simulation. The rejection probability of our solution is notably higher than the fair share one. As already stated, this is because of the stringent requirement enforced by our solution in order to guarantee the zero rebuffering time for every accepted client. In other words, warranting high-quality video experience ( 5 Mpps) with a stall-free session is done at the expense of over provisioning the bandwidth resource. Moreover, network calculus considers the worst-case analysis and sometimes its assumptions are too pessimistic as reported in previous works.

*3) Rebuffering Duration:* One of the most crucial factors in measuring user satisfaction is the number of rebuffering events and for how long they occur. As shown in Fig. 7, the accepted clients in our solution do not encounter any rebuffering events during the whole session. This is thanks to the strict formula (10) used to measure the chunks' delay and the double check process (see Algorithm 1, line 19 and its explanation in Section IV-B) that ensures there is no impact from the newly connected client on the maximum delay of the previously accepted clients. Contrary, the fair share solution accepts as many clients as possible, and fairly share the selected paths between the interfering flows by decreasing the video representation of the clients playing at high levels. This strategy resulted in an over-subscribed network that potentially leads to chunks' deadline misses and rebuffering events as depicted in Fig. 7.

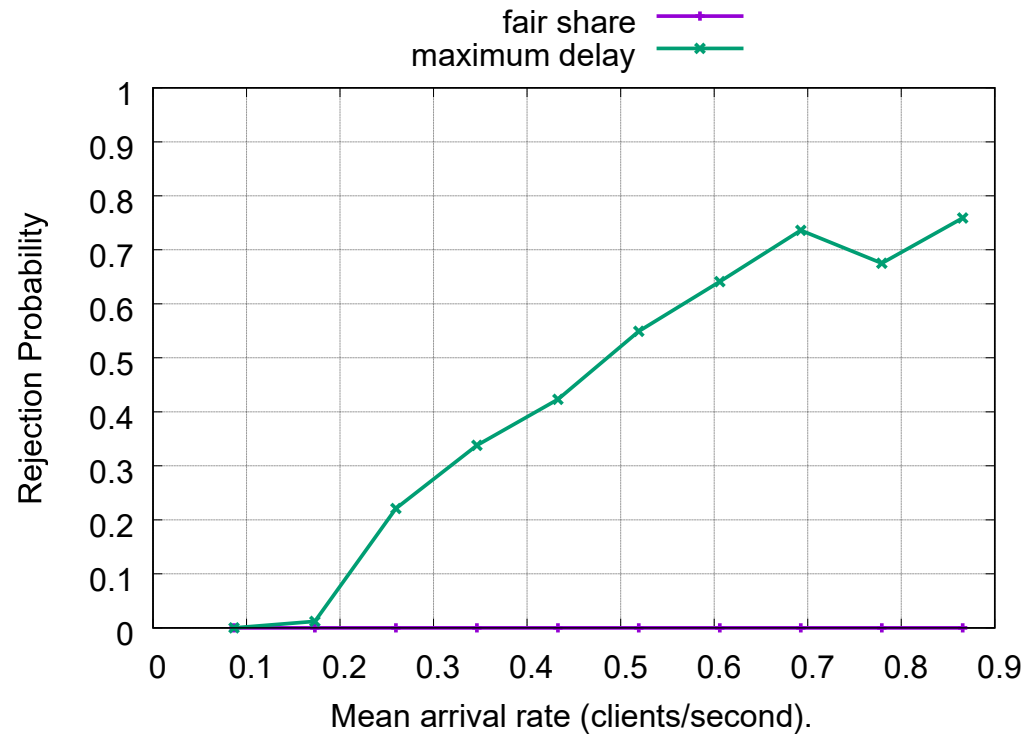

Fig. 6: Rejection Probability.

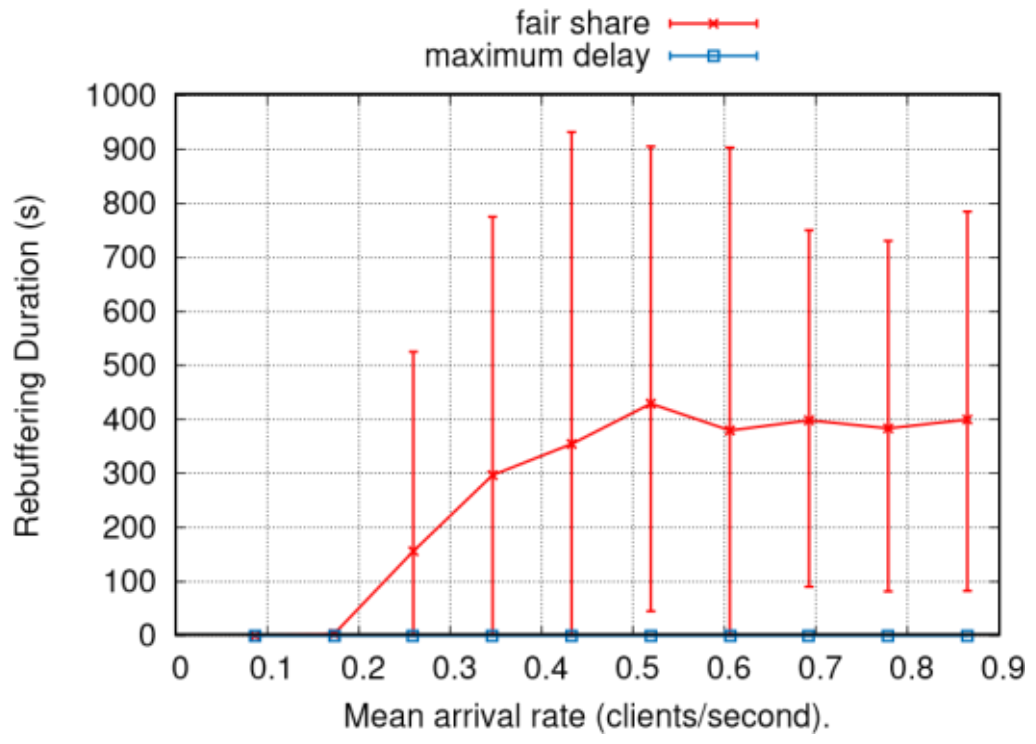

Fig. 7: Cumulative rebuffering duration.

## VI. Conclusion

Ensuring high user QoE in media services remains a challenging task. In this paper, we have proposed a new DASHbased solution using deterministic network calculus theory in SDN-enabled networks to provide users with high video quality while ensuring a stall-free session in order to provide the user with high QoE. The proposed solution leverages the benefits of SDN to provide the clients with the highest possible video quality that guarantees a maximum delay lower than the chunk size. This results in a stall-free session which highly improves the user's QoE. As the proposed solution over-provision the network to guarantee a maximum delay for the video chunks, this makes the proposed solution highly important in scenarios where missing the chunks' deadlines is impermissible such as remote control/driving of an autonomous vehicle and serving video premium subscribers in future 5G mobile networks.

To further improve the proposed model and in order to reduce the number of rejected clients, we plan to adjust the current model and find the tightest possible upper-bound that ensures a maximum end-to-end delay which is lower than the chunk size.

### Acknowledgement

This work was partially supported by the European Union's Horizon 2020 research and innovation programme under the MATILDA project with grant agreement No. 761898, and by the Academy of Finland's Flagship programme 6Genesis under grant agreement No. 318927. It was also supported in part by the Academy of Finland under CSN project with grant No. 311654.